\documentclass[10pt]{article}
\usepackage{amsfonts,latexsym}

\catcode`\@=11

\newcommand{\be}[1]{\begin{equation}\label{#1}}
\newcommand{\ee}{\end{equation}}
\newcommand{\ba}[1]{\begin{eqnarray}\label{#1}}
\newcommand{\ea}{\end{eqnarray}}
\newcommand{\rf}[1]{(\ref{#1})}
\newcommand{\nn}{\nonumber}
\newcommand{\bmatrix}[1]{\left( \begin{array}{#1}}
\newcommand{\ematrix}{\end{array}\right)}
\newcommand{\opensquare}{\mbox{$\rlap{$\sqcap$}\sqcup$}}
\newcommand{\diag}{\mbox{\rm diag}}

\font\msbm=msbm10
\def\RR{\hbox{\msbm R}}

\begin{document}

\author{U.G\"unther\dag\footnote{e-mail: u.guenther@htw-zittau.de},
A.Zhuk\ddag\S\footnote{e-mail: zhuk@paco.odessa.ua}\\[3ex]
\dag Gravitationsprojekt, Mathematische Physik I,\\
Institut f\"ur Mathematik,
Universit\"at Potsdam,\\
Am Neuen Palais 10, PF 601553, D-14415 Potsdam, Germany\\[1ex]
\ddag Max-Planck-Institut f\"ur Gravitationsphysik, 
Albert-Einstein-Institut,\\ 
Schlaatzweg 1, D-14473 Potsdam, Germany\\[1ex]
\S Department of Physics, University of Odessa,\\ 
2 Petra Velikogo St., Odessa 270100, Ukraine}
\title{\bf Observable effects from extra dimensions}

\date{29.05.1999}

\maketitle

\abstract{
{} For any multidimensional theory with  compactified internal spaces,
conformal excitations of the internal space metric result in gravitational
excitons in the external spacetime. These excitations contribute either 
to dark matter
or
to  cross sections of usual particles.
}

\vspace*{.5cm}

%
The large-scale dynamics of the observable part of our present time 
universe is well described by the Friedmann model with four-dimensional 
Friedmann-Robertson-Walker metric. However, it is possible that spacetime 
at short (Planck) distances might have a dimensionality of more than four 
and possess a rather complex topology. String theory and its recent 
generalizations --- $p$-brane, $M$- and $F$-theory widely use this concept
and give it a new foundation.
Most of these unified models
are initially  constructed on a higher-dimensional spacetime manifold,
say of dimension $D>4$, which then
undergoes some scheme of spontaneous compactification
yielding a direct product manifold $M^4\times K^{D-4}$
where $M^4$ is the manifold of spacetime
and $K^{D-4}$ is a compact internal space. Hence it is natural to investigate
observable consequences of such a compactification hypothesis.

One of the main problems in multidimensional models consists in the dynamical 
process leading from a stage with all dimensions developing on the same scale 
to the actual stage of the universe, where we have only four external 
dimensions and all internal spaces have to be compactified and contracted 
to sufficiently small scales, so that they are apparently unobservable.
To make the internal dimensions unobservable at the actual stage of the
universe we have to demand their contraction to scales $10^{-17}cm$ ---
$10^{-33}cm$ (between the Fermi and Planck lengths). This leads to 
an effectively four-dimensional universe. 
However, there is still a question on possible observable phenomena
following from such small compactified spaces.
In the present paper we predict some physical effects which should take
place in this case.

{}As starting point let us consider a simple multidimensional cosmological
model (MCM) with spacetime manifold  
\begin{equation}
\label{2.1}M=M_0\times M_1\times \dots \times M_n
\end{equation}
and with decomposed metric on $M$
\begin{equation}
\label{2.2}g=g_{MN}(X)dX^M\otimes dX^N=g^{(0)}+\sum_{i=1}^ne^{2\beta
^i(x)}g^{(i)},
\end{equation}
where $x$ are some coordinates of the $D_0=d_0+1$ - dimensional manifold $%
M_0 $ and
\begin{equation}
\label{2.3}g^{(0)}=g_{\mu \nu }^{(0)}(x)dx^\mu \otimes dx^\nu .
\end{equation}
Let the manifolds $M_i$ be $d_i$-dimensional Einstein spaces with metric $%
g^{(i)}=g^{(i)}_{m_in_i}(y_i)dy_i^{m_i}\otimes dy_i^{n_i},$ i.e.,
\be{2.4}
R_{m_in_i}\left[ g^{(i)}\right] =\lambda
^ig_{m_in_i}^{(i)},\qquad m_i,n_i=1,\ldots ,d_i \qquad
\mbox{and} \quad
R\left[ g^{(i)}\right] =\lambda ^id_i\equiv R_i.
\ee
In the case of constant curvature spaces parameters $\lambda ^i$ are
normalized as $\lambda ^i=k_i(d_i-1)$ with $k_i=\pm 1,0$. The internal spaces $%
M_i\quad (i=1,\dots ,n)$ may have nontrivial global topology, being compact
(i.e. closed and bounded) for any sign of spatial topology.

With total dimension $D=1+\sum_{i=0}^nd_i$, $\kappa ^2$ a $D$-dimensional
gravitational constant, $\Lambda $ - a $D$-dimensional bare cosmological
constant we consider an action of the form
\begin{equation}
\label{2.6}S=\frac 1{2\kappa ^2}\int\limits_Md^DX\sqrt{|g|}\left\{
R[g]-2\Lambda \right\} .
\end{equation}
Of course, the ansatz of our model with this action 
is rather simplified and can describe only partial aspects 
of a more realistic theory. The $\Lambda$ - term can originate, for example, 
from $D-1$ - form gauge fields \cite{W}.
We also could  enlarge the model action e.g. by inclusion of a dilatonic
scalar field
as well as other matter fields, take into account the Casimir effect
due to non-trivial topology of the  manifold \rf{2.1}, 
consider different monopole ans\"atze etc. However, to reveal observable
effects following from extra dimensions  
it is sufficient to consider such a simplified model.

After dimensional reduction and a conformal transformation to the
Einstein frame
\be{2.10}
g_{\mu \nu }^{(0)}=\Omega ^2\tilde g_{\mu \nu }^{(0)}
=exp\left( -\frac 2{D_0-2}\sum_{i=1}^nd_i\beta ^i\right)
\tilde g_{\mu \nu }^{(0)}
\ee
the action reads \cite{GZ}
\begin{equation}
\label{2.12}S=\frac 1{2\kappa _0^2}\int\limits_{M_0}d^{D_0}x\sqrt{|\tilde
g^{(0)}|}\left\{ \tilde R\left[ \tilde g^{(0)}\right] -\bar G_{ij}\tilde
g^{(0)\mu \nu }\partial _\mu \beta ^i\,\partial _\nu \beta
^j-2U_{eff}\right\} ,
\end{equation}
where $\kappa _0^2=\kappa ^2/V_I$ is the $D_0$-dimensional gravitational
constant and
$V_I=\prod_{i=1}^nv_i=\prod_{i=1}^n\int_{M_i}d^{d_i}y\sqrt{%
|g^{(i)}|}$
defines the internal space volume corresponding to the scale factors $%
a_i\equiv 1,i=1,\dots ,n$. 
The tensor components of the midisuperspace metric (target space metric on $%
\RR _T^n$) $\bar G_{ij}\ (i,j=1,\ldots ,n)$
and the effective potential are respectively
\begin{equation}
\label{2.13}\bar G_{ij}=d_i\delta _{ij}+\frac 1{D_0-2}d_id_j
\end{equation}
and
\begin{equation}
\label{2.15}U_{eff}={\left( \prod_{i=1}^ne^{d_i\beta ^i}\right) }^{-\frac
2{D_0-2}}\left[ -\frac 12\sum_{i=1}^nR_ie^{-2\beta ^i}+\Lambda +\kappa
^2\rho \right] ,
\end{equation}
where the phenomenological energy density $\kappa ^2\rho $ 
is not specified here. Depending on the concret model \cite{GZ} 
it takes into account e.g. the Casimir effect of additional matter fields or
Freund-Rubin monopoles. 
In the case of a purely geometrical model it vanishes.

Variation of action \rf{2.12} with respect to $\tilde g^{(0)}$ and $\beta^i$
gives the Euler-Lagrange
equations for the scale factors and the external metric.
Assuming that there exists a well defined splitting of
the physical fields $(\tilde g^{(0)},\beta )$ into not necessarily constant
background
components $(\bar g,\bar \beta )$ and small perturbational (fluctuation)
components $(h,\eta )$
\ba{4g}
\tilde g^{(0)}_{\mu \nu} & = & \bar g_{\mu \nu} + h_{\mu \nu}\, ,
\\\nn
\beta ^i & =& \bar \beta ^i+ \eta ^i
\ea
we can perform a perturbational analysis of the interaction
dynamics. For example, for the internal space scale factors we obtain 
in zeroth and first order approximation respectively \cite{gsz1}
\ba{5g2}
& &\opensquare  \bar \beta ^i =
\left[\bar G^{-1}\right]^{ij}b_j(\bar \beta )\, , \\
& &\opensquare \eta^i - \left[\bar G^{-1}\right]^{ij}
A_{jk}(\bar \beta )\eta ^k
=\frac{1}{\sqrt{|\bar g|}}\partial_{\nu}
\left(\sqrt{|\bar g|}h^{\mu \nu}\partial_{\mu}\bar \beta^i\right) -
\frac12\bar g^{\mu\nu}\partial_{\mu}\bar \beta^i\partial_{\nu}h \label{7g} \ ,
\ea
where we abbreviate
\begin{equation}
\label{g1a}A_{ij}:=\frac{\partial ^2U_{eff}}{\partial \beta ^i\partial
\beta ^j},\quad b_i:=\frac{\partial U_{eff}}{\partial \beta ^i}\ .
\end{equation}
In equations \rf{5g2} and \rf{7g}
the covariant derivative is taken with respect to the background metric
$\bar g_{\mu \nu }$ and indices in $h_{\mu \nu }$ are raised and lowered 
by the background metric
$\bar g_{\mu \nu}$, e.g. $h =h_{\mu \nu}\bar g^{\mu \nu}$.

We can diagonalize matrix 
$\left[ \bar G^{-1}A\right]
_k^i\equiv \left[ \bar G^{-1}\right] ^{ij}A_{jk}(\bar \beta )$
by an appropriate background depending $SO(n)$ - rotation $S=S(\bar \beta )$
\be{8g}
S^{-1}\bar G^{-1}AS\stackrel{def}{=}M^2=\diag\left[
m_1^2(\bar \beta ),\ldots  ,m_n^2(\bar \beta )\right] 
\ee
and rewrite Eq. (\ref{7g}) in terms of generalized normal modes
(gravitational excitons \cite{GZ})
$\psi =S^{-1}\eta $:
\be{9g}
\bar g^{\mu \nu }D_{\mu}D_{\nu}\psi - M^2(\bar \beta )\psi
=\left(h^{\mu \nu }-\frac12\bar g^{\mu \nu }h\right)_{;\nu }D_{\mu }
\bar \varphi
+h^{\mu \nu }D_{\mu }D_{\nu }\bar \varphi \ ,
\ee
where $\bar \varphi $ are $SO(n)-$rotated background scale factors
$\bar \varphi = S^{-1}\bar \beta $ and $M^2$ can be interpreted as
background depending diagonal mass matrix for the gravitational excitons.

$D_\mu $ denotes
a covariant derivative 
\begin{equation}
\label{10g}D_\mu :=\partial _\mu +\Gamma _\mu +\omega _\mu \ ,
\quad \omega _\mu
:=S^{-1}\partial _\mu S
\end{equation}
with $\Gamma _\mu +\omega _\mu $ as connection
on the fibre bundle $E(M_0,\RR^{D_0}\oplus \RR^n_T) \to M_0$
consisting of the base
manifold $M_0$ and vector spaces $\RR_x^{D_0}\oplus \RR_{Tx}^n=T_x
M_0\oplus \left\{ \left( \eta ^1(x),\ldots ,\eta ^n(x)\right) \right\} $
as fibres.   So, the background components $\bar \beta ^i(x)$
via the effective potential $U_{eff}$ and its Hessian $A_{ij}(\bar \beta)$
play the role
of a medium for the gravitational excitons $\psi ^i(x).$ Propagating in 
$M_0$ filled with this medium  they change their masses
as well as the direction of
their "polarization"
defined by the unit vector in the fibre space 
\begin{equation}
\label{g11}\xi (x):=\frac{\psi (x)}{\left| \psi (x)\right| }\in
S^{n-1}\subset \RR^n,
\end{equation}
where $S^{n-1}$ denotes the $(n-1)-$dimensional sphere.

{}From (\ref{7g}) and (\ref{9g}) we see that
in the lowest order (linear) approximation of the used
perturbation theory
a non-constant scale factor background
is needed for an
interaction between gravitational excitons and gravitons.
This can be also easily seen from the interaction term
in the action functional (in the lowest order approximation and in traceless
gauge: $h = 0 $)
\be{20g}
S_{int} = \frac{1}{\kappa _0^2}\int \limits_{M_0}d^{D_0}x \sqrt
{|\bar g|} h^{\mu \nu}\bar G_{ij}\partial_{\mu}\bar \beta^i
\partial_{\nu}\eta ^j \, .
\ee

{}For constant scale factor backgrounds $\bar \beta =const$ the
system is necessarily located in
one of the minima $\bar \beta =\beta _{(c)}$ of the effective
potential $U_{eff}$ so that $b_i(\beta _{(c)})=0$, and 
$U_{eff}(\beta _{(c)})=\Lambda _{(c)eff}$ 
 plays the role of a $D_0$-dimensional effective
cosmological constant (according to recent observational data there is a 
strong evidence for a positive cosmological constant of the universe
 \cite{PTW}).
Gravitational excitons and gravitons can in this case only interact
via nonlinear (higher order) terms. In the linear approximation they
decouple over constant scale factor backgrounds due to vanishing terms in
(\ref{7g}) and (\ref{9g}) (see also \rf{20g}). This means that in this case 
conformal excitations of the metric of the internal spaces behave as
non-interacting
massive scalar particles developping on the background of the external
spacetime. Due to their vanishing cross section
they will contribute only to dark matter.

{}From the geometrical point of view
it is clear that gravitational excitons are an
inevitable consequence of the existence
of extra dimensions. For any theory with  compactified internal spaces
conformal excitations of the internal space metric will result
in gravitational
excitons in the external spacetime. The form of the effective potential
as well as
masses of gravitational excitons and the value of the effective cosmological 
constant are model dependent.
It is important to note that even for internal spaces compactified at 
Planck scales
the masses
of gravitational excitons can run a very wide range of values --- from
very heavy (of order of the Planck mass) to extremely light.
It depends on  the parameters of a the concrete model. For example, 
models with one internal factor space 
and nonvanishing energy density $\kappa ^2\rho \neq 0$ induced e.g. 
by Freund-Rubin monopoles or the Casimir effect of additional matter fields, 
or with vanishing $\kappa ^2\rho \equiv 0$,  yield exciton masses which are up to 
a numerical prefactor of order one
\cite{GZ}
\be{19}
m \sim \left( a^{-1}_{(c)}\right)^{\frac{D-2}{2}} \, .
\ee
Here we assumed $D_0 = 4$, and $a_{(c)}$ is the compactification scale 
of the internal factor space. 
This means  that if compactification takes place
at $a_{(c)} = 10^2 L_{Pl}$ then $m \sim 10^{-8}m_{Pl}$ and 
$m \sim 10^{-24}m_{Pl}$ for $D = 10$ and $D = 26$ correspondingly.  
Of course, gravitational excitons can be excited at the present 
time if their masses are much less than the Planck mass. 
So, even for compactification scales 
very close to the Planckian one, masses of the 
gravitational excitons can correspond to energies which are achieved 
by present accelerators.

On the other hand, in the case of a
non-constant internal scale factor background
gravitational excitons interact with usual matter already in the lowest
order approximation. If such interactions are strong enough then gravitational 
excitons cannot be considerad as dark matter and they should 
contribute to the cross sections of usual particles. Equations \rf{9g}
and \rf{20g} show for example that gravitational excitons can produce
gravitons and vice versa. The form of interactions will depend on the
concrete model.
This should give a possibility to check experimentally
different models on their compatibility with observational data.
Possible interaction channels for tests could be e.g. 
interactions between gravitational
excitons and abelian gauge fields or gravitational excitons
and spinor fields \cite{gsz1}.

Other interesting physical effects can be expected in the  vicinity of 
topologically non-trivial objects such as black holes or cosmic strings, 
and in a
multiply connected universe, e.g. in a universe with 
lorentzian wormholes (if they connect different regions of the same universe)
or in a universe with a compact space manifold  with negative 
or zero constant curvature. 
Propagating from a source to an observer on different sides
of the topologically
non-trival object gravitational excitons can via $SO(n)-$rotation
of the polarization vector in the target space accquire different
polarizations (similar to $SO(2)\approx U(1)-$Aharonov-Bohm---phase-rotations 
in QED). In the observation region this will result
in a local interference. Allowing the gravitational excitons to interact
with other fields the interference picture should be observable.

\mbox{} \\
{\bf Acknowledgements}\\
The work  was partially supported
by the Max-Planck-Institut f\"ur Gravitationsphysik (A.Z.) and DFG grant
436 UKR 113 (U.G.).
 
A.Z. thanks Professor Nicolai and the
Albert-Einstein-Institut f\"ur Gravitationsphysik
for their hospitality while this paper was written.

\end{document}